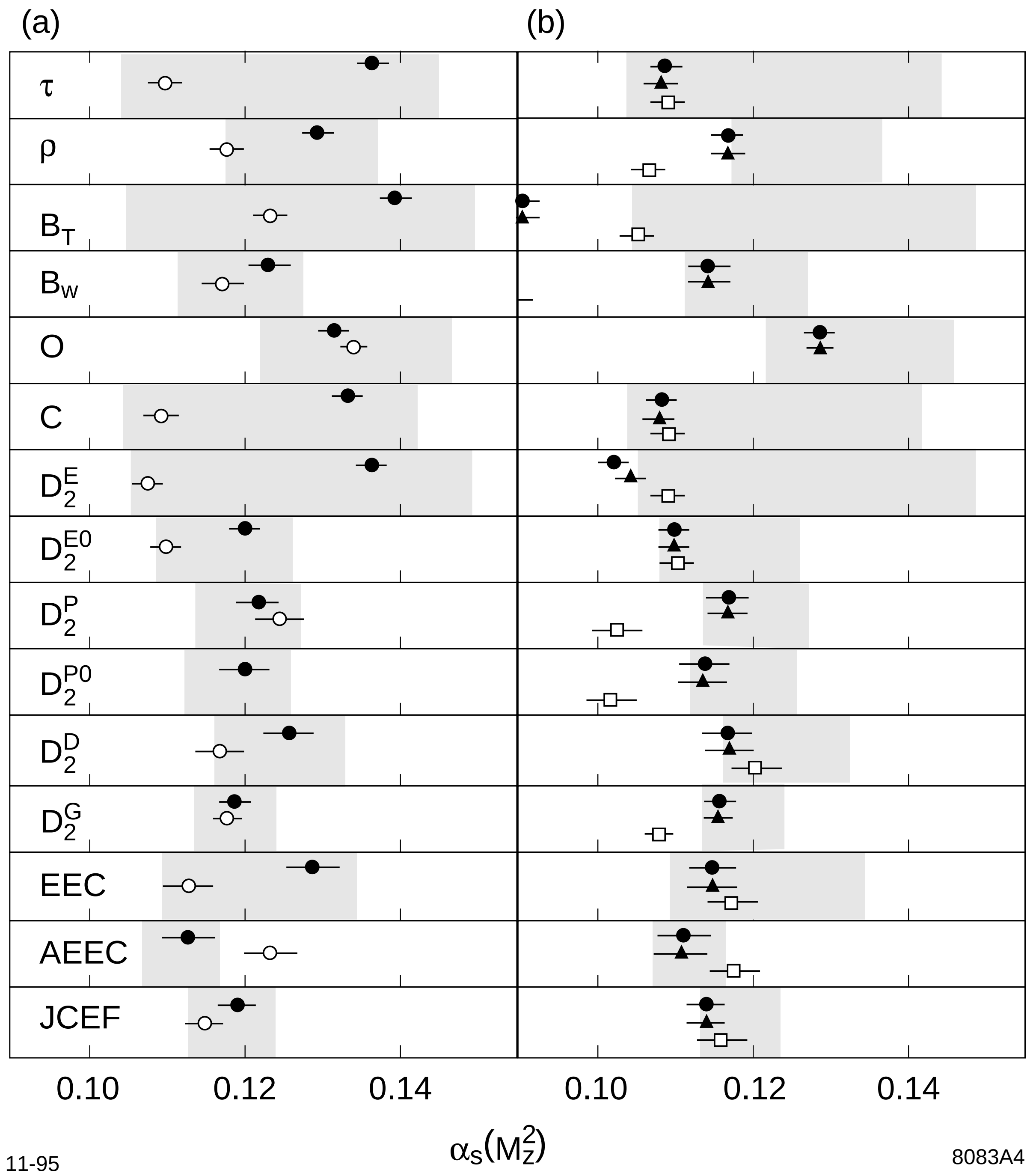

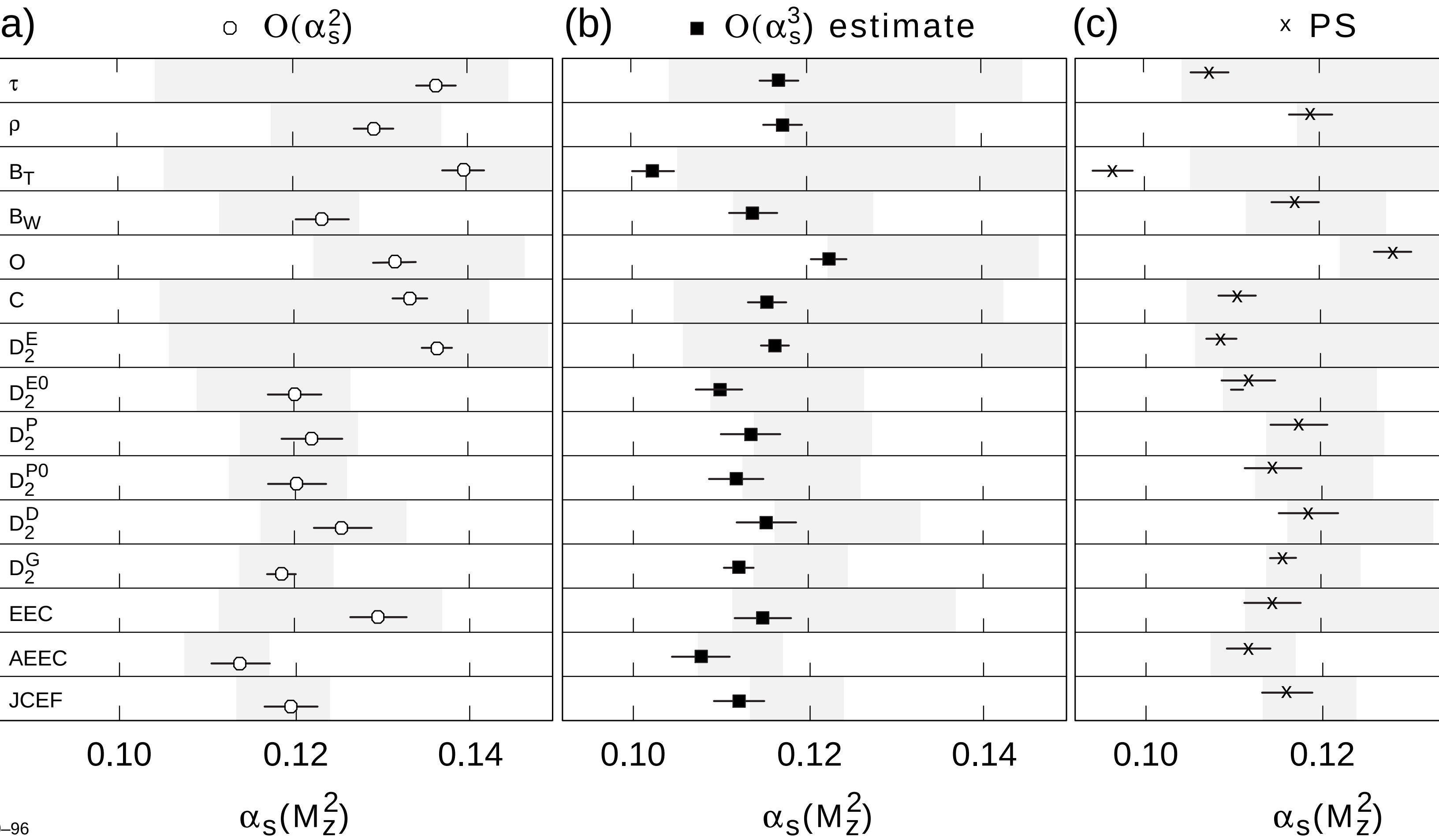

# DETERMINATION OF $\alpha_s(M_Z^2)$ FROM HADRONIC EVENT SHAPE OBSERVABLES IN $e^+e^-$ ANNIHILATION*

P.N. Burrows**

*Stanford Linear Accelerator Center*
*Stanford University, Stanford, CA94309, USA*

Email: burrows@slac.stanford.edu

## ABSTRACT

The determination of $\alpha_s(M_Z^2)$ using $O(\alpha_s^2)$ calculations of hadronic event shape observables in $e^+e^-$ annihilation is reviewed. The large scatter among $\alpha_s(M_Z^2)$ values determined from different observables may be interpreted as arising from the effect of uncalculated higher-order contributions. The application of 'optimised' perturbation theory and Padé approximants in an attempt to reduce this effect is discussed.

*Talk presented at the*
*28th International Conference on High Energy Physics*
*25-31 July 1996, Warsaw, Poland*

**Permanent address: Laboratory for Nuclear Science, Massachusetts Institute of Technology, Cambridge, MA 02139, USA.

* Work supported by Department of Energy contracts DE–AC02–76ER03069 (MIT) and DE–AC03–76SF00515 (SLAC).

One of the most important tasks in high energy physics is the precise determination of the strong coupling $\alpha_s(M_Z^2)$. Not only does measurement of $\alpha_s(M_Z^2)$ in different hard processes and at different hard scales $Q$ provide a fundamental test of the theory of strong interactions, Quantum Chromodynamics (QCD), but it also allows constraints on extensions to the Standard Model of elementary particles [1]. The large set of $\alpha_s(M_Z^2)$ measurements is consistent with a central value of about 0.117 with an uncertainty of $\pm 0.005$ [2]. However, nearly all measurements are limited by theoretical uncertainties that derive from lack of knowledge of higher-order perturbative QCD contributions, or of non-perturbative effects, or both. It is hence vital to reduce the size of the limiting theoretical uncertainties which may, or may not, be concealing new physics.

Here hadronic event shape observables in $e^+e^-$ annihilation are considered. For an infra-red- and collinear-safe observable $X$:

$$\frac{1}{\sigma}\frac{d\sigma}{dX}(X,\mu) = \overline{\alpha_s}(\mu)A(X) + \overline{\alpha_s}^2(\mu)\ B(X,\mu) + \overline{\alpha_s}^3(\mu)\ C(X,\mu) + O(\overline{\alpha_s}^4(\mu)) \qquad (1)$$

where $\overline{\alpha_s} \equiv \alpha_s/2\pi$ and $\mu$ is the renormalisation scale. To date only the coefficients $A(X)$ and $B(X,\mu)$ have been calculated [3,4]. The 15 hadronic event shape observables used in the recent $\alpha_s(M_Z^2)$ determination by the SLD Collaboration [5] were employed. Distributions of these observables were measured [5] using a sample of about 50,000 hadronic $Z^0$ decay events. The data were corrected for detector bias effects such as acceptance, resolution, and inefficiency, as well as for the effects of initial-state radiation and hadronisation, to yield 'parton-level' distributions, which can be compared directly with the QCD calculations. The EVENT program [6] was used to calculate the coefficients $A$ and $B$ in Eq. (1).

First, the $O(\alpha_s^2)$ calculation, using the *physical scale* $\mu = M_Z$, was fitted [7] to the measured parton-level distributions by minimising $\chi^2$ w.r.t. variation of $\Lambda_{\overline{MS}}$. Each resulting $\Lambda_{\overline{MS}}$ value was translated into $\alpha_s(M_Z^2)$ and is shown, with experimental errors [5], in Fig. 1(a). There is considerable scatter among the 15 $\alpha_s(M_Z^2)$ values. Similar results have been observed previously [8]. Since the same data sample was used to measure each observable, and since the observables are highly correlated, this scatter is very significant and can be interpreted as arising from uncalculated higher-order perturbative QCD contributions, which *a priori* may be of different sign and magnitude for the different observables. The average $\alpha_s(M_Z^2)$ value and corresponding r.m.s. deviation are listed in Table 1.

This procedure was repeated [7] using the *experimentally-optimised-scale* approach [9] in which a simultaneous fit of $\Lambda_{\overline{MS}}$ and $\mu$ to each observable was performed. Each resulting $\Lambda_{\overline{MS}}$ value was translated to $\alpha_s(M_Z^2)$ and is shown in Fig. 1(a). For $D_2^{P0}$ no minimum in $\chi^2$ w.r.t. variation of $\mu$ in the range $10^{-4} \leq \mu^2/M_Z^2 \leq 10^2$ could be found [7]. Again, there is large scatter among the 14 $\alpha_s(M_Z^2)$ values. For most observables the experimentally-optimised scale yields a lower value of $\alpha_s(M_Z^2)$ than the physical scale because the optimised scale is typically smaller than $M_Z$, requiring a smaller value of $\Lambda_{\overline{MS}}$ in order to fit the data [10]. The average $\alpha_s(M_Z^2)$ value and r.m.s. deviation are listed in Table 1. The r.m.s. deviation is comparable with that resulting from the choice of the physical scale, implying that use of the experimentally-optimised scale does not serve to reduce uncalculated higher-order effects.

From the $\mu$-dependence a 'renormalisation scale uncertainty' on $\alpha_s(M_Z^2)$ can be defined [5] for each observable; these are shown as bands in Figs. 1,2. Within such uncertainties the $\alpha_s(M_Z^2)$ values determined from the different observables using either the physical or experimentally-optimised scales are consistent, but this arbitrary procedure leads to a large uncertainty of $\pm 0.0106$ on the average value of $\alpha_s(M_Z^2)$ [5].

The best resolution of this situation would be to calculate the observables to higher order in perturbation theory, a difficult and unattractive task that has not yet been achieved. In the absence of $O(\alpha_s^3)$ QCD calculations it has been suggested that the $O(\alpha_s^2)$ calculations can be 'optimised' by choosing a specific value of the renormalisation scale. Since the all-orders result would be independent of renormalisation scale, Stevenson suggests that $\mu$ be chosen according to the 'Principle of Minimal Sensitivity' (PMS) [11], so that $\partial\sigma(X,\mu)/\partial\mu = 0$. Grunberg suggests that $\mu$ be chosen to give the 'fastest apparent convergence' (FAC) of the series [12], so that the second-order term in Eq. (1) vanishes. Brodsky, Lepage and Mackenzie advocate that $\mu$ be chosen to remove the $N_f$-dependence of the second-order term in Eq. (1), effectively incorporating quark and gluon vacuum polarisation contributions into the definition of the strong coupling [13].

For each observable the PMS, FAC and BLM optimised scales were calculated [7] and used in turn in a fit of the $O(\alpha_s^2)$ calculation to each measured distribution to determine $\Lambda_{\overline{MS}}$ and hence $\alpha_s(M_Z^2)$. The results are shown in Fig. 1(b); in the case of the oblateness $O$ an acceptable fit with the BLM scale could not be obtained. For each observable the PMS- and FAC-derived $\alpha_s(M_Z^2)$ values are very similar, whereas, in some cases, the BLM-derived $\alpha_s(M_Z^2)$ value differs from them. This behaviour follows from the correlation between the scale value and the corresponding $\Lambda_{\overline{MS}}$ required to fit the data [10]. For a given observable the PMS- and FAC-derived $\alpha_s(M_Z^2)$ values are often, though not always, close to that determined using the experimentally-optimised scale. Furthermore, for most observables the PMS-, FAC- and BLM-derived $\alpha_s(M_Z^2)$ values all lie within the range encompassed by the renormalisation scale uncertainty defined in Ref. [5], though for $\rho$, $B_W$, $D_2^P$, $D_2^{P0}$, $D_2^G$ and $(B_T)$, the BLM- (PMS/FAC-) derived values lie below this range.

For any of the PMS, FAC or BLM scale choices there is considerable scatter among the $\alpha_s(M_Z^2)$ values from all the observables. In each case the average over the $\alpha_s(M_Z^2)$ values, and corresponding r.m.s. deviation, are shown in Table 1. The r.m.s. deviations are comparable with those resulting from choice of the physical and experimentally-optimised scales, implying that higher-order effects contribute roughly equally in all of these procedures.

An approach for estimating higher-order perturbative contributions to, as well as the sum of, perturbative QCD series is based on Padé Approximants (PA). The PA $[N/M]$ to the series:

$$S \quad = \quad S_0 \; + \; S_1 x \; + \; S_2 x^2 \; + \; \ldots \; + \; S_{N+M} x^{N+M} \tag{2}$$

is defined [14]:

$$[N/M] \quad \equiv \quad \frac{a_0 + a_1 x + a_2 x^2 + \ldots + a_N x^N}{1 + b_1 x + b_2 x^2 + \ldots + b_M x^M}, \tag{3}$$

where $N$ and $M$ are integers such that $N \geq 0$ and $M > 0$, and

$$[N/M] \quad = \quad S + O(x^{N+M+1}). \tag{4}$$

The coefficients $a_i$ $(0 \leq i \leq N)$ and $b_j$ $(1 \leq j \leq M)$ are obtained by multiplying Eq. 4 by the denominator of Eq. 3 and equating coefficients of like powers of $x$. By consideration of the terms of O$(x^{N+M+1})$ one can obtain an estimate of the coefficient $S_{N+M+1}$. Furthermore, for an asymptotic series $[N/M]$ can be taken to be an estimate of the sum (PS) of the series to all orders.

In the case of hadronic event shape observables the PA [0/1] can be defined for the series Eq. (1) and, for each bin of each observable, was used to derive an estimate of the coefficient $C$ of the O($\alpha_s^3$) term [15]. The PA prediction for $C$ was added to the exact O($\alpha_s^2$) calculation to obtain an estimate of the series to O($\alpha_s^3$). For each observable the calculation was fitted to the data [5] using $\mu = M_Z$, and the resulting $\alpha_s(M_Z^2)$ values are shown in Fig. 2(a). The O($\alpha_s^3$) estimate does not provide a good fit to the $B_T$ data [15] and this observable is excluded from further discussion. In each case the $\alpha_s(M_Z^2)$ value derived using the O($\alpha_s^3$) estimate is lower than that derived using the O($\alpha_s^2$) calculation, which is expected since $C$ is positive [15], and the O($\alpha_s^3$) $\alpha_s(M_Z^2)$ value lies near the lower bound given by the scale uncertainty on the O($\alpha_s^2$) result. To the extent that the PA O($\alpha_s^3$) estimate is accurate, this implies that the renormalisation scale uncertainty assigned to the O($\alpha_s^2$) $\alpha_s(M_Z^2)$ value from each observable is a reasonable estimate of the effect of the missing O($\alpha_s^3$) contribution. The average and r.m.s. deviation of the 14 $\alpha_s(M_Z^2)$ values are listed in Table 1. The scatter is noticeably smaller than in any of the O($\alpha_s^2$) cases, implying that the Padé method provides at least a partial approximation of higher-order perturbative QCD contributions to event shape observables.

Finally, the PS [0/1] was used as an estimate of the sum of the asymptotic series and $\alpha_s(M_Z^2)$ was extracted by comparison with the data in a similar manner [15]. The $\alpha_s(M_Z^2)$ values are shown in Fig. 2(b). Typically, for each observable, the PS $\alpha_s(M_Z^2)$ value is close to the PA O($\alpha_s^3$) value. Again the fit to $B_T$ is very poor [15]. The average and r.m.s. deviation over the set of 14 $\alpha_s(M_Z^2)$ values are listed in Table 1. Though the average value is close to that obtained using the PA O($\alpha_s^3$) estimate, the r.m.s. deviation is somewhat larger, implying that the PS [0/1] provides a poorer estimate of the sum of the series than the PA [0/1] estimate to O($\alpha_s^3$).

In summary, $\alpha_s(M_Z^2)$ has been determined by fitting $O(\alpha_s^2)$ QCD predictions of 15 hadronic event shape observables to $e^+e^-$ annihilation data at the $Z^0$ resonance collected by the SLD experiment. Five prescriptions for choosing the renormalisation scale were used, namely the physical, experimentally-optimised, PMS-, FAC- and BLM-optimised scales. Though the average $\alpha_s(M_Z^2)$ value, taken over all the observables, differs among these five procedures, the scatter among the $\alpha_s(M_Z^2)$ values from different observables is equally large in each case, the r.m.s. deviation being about 0.008, implying that these specific renormalisation scale choices do not offer any numerical advantage in terms of the accuracy of $O(\alpha_s^2)$ perturbative QCD predictions of $e^+e^-$ event shapes.

If Padé Approximants are used to estimate the $O(\alpha_s^3)$ terms the scatter among the $\alpha_s(M_Z^2)$ values from different observables is reduced to $\pm 0.0035$. This is comparable with the combined experimental error and hadronisation uncertainty on a single observable measured at $Q = M_Z$ [5]. Since the accuracy of the Padé Approximant method can only be verified *a posteriori*, exact calculation of the $O(\alpha_s^3)$ terms in order to confirm these results is extremely desirable.

| Procedure | $\alpha_s(M_Z^2)$ |
|---|---|
| Physical scale | $0.1265 \pm 0.0076$ |
| Exp. opt. scale | $0.1173 \pm 0.0071$ |
| PMS scale | $0.1123 \pm 0.0079$ |
| FAC scale | $0.1123 \pm 0.0080$ |
| BLM scale | $0.1088 \pm 0.0075$ |
| Padé O($\alpha_s^3$) | $0.1147 \pm 0.0035$ |
| Padé sum | $0.1148 \pm 0.0052$ |

Table 1: Mean and r.m.s. $\alpha_s(M_Z^2)$ values determined using different theoretical procedures.

**Figure Captions**

FIG. 1. Values of $\alpha_s(M_Z^2)$ from QCD fits to the data using: (a) physical (solid circles), and experimentally-optimised (open circles) scales; (b) PMS- (solid circles), FAC- (solid triangles), and BLM- (open squares) optimised scales. In all cases only experimental error bars are shown. For each observable the shaded region indicates the total uncertainty estimated in Ref. [5], dominated by the contribution from wide variation of the renormalisation scale.

FIG. 2. Values of $\alpha_s(M_Z^2)$ from QCD fits to the data using: (a) PA O($\alpha_s^3$) estimate (squares); (b) Padé sum (PS) (crosses). In all cases only experimental error bars are shown. For each observable the shaded region indicates the total uncertainty estimated in Ref. [5], dominated by the contribution from wide variation of the renormalisation scale.